\documentclass[11pt,a4paper]{article}
\pdfoutput=1
\usepackage{jheppub}
\usepackage{color,array,subfigure,slashed,wasysym}
 \usepackage[default]{dhucs-interword}
\usepackage[hangul]{dhucs-setspace}
\usepackage{multirow, graphicx,url,mathrsfs}
\usepackage{wrapfig,boxedminipage,setspace,subfigure,epsfig}
\usepackage{amsxtra,amstext,latexsym,dsfont}
\usepackage{amsmath,amssymb,amsbsy,amsfonts,latexsym,hyperref}
\usepackage{cancel}

\usepackage{multirow, graphicx,amssymb,url,mathrsfs,amsmath}
\usepackage{wrapfig,boxedminipage,setspace,subfigure,epsfig}
\usepackage{amsxtra,amstext,latexsym,dsfont,amsfonts}
\usepackage{color,eucal}

\newcommand{\be}{\begin{equation}}
\newcommand{\ee}{\end{equation}}
\newcommand{\bea}{\begin{eqnarray}}
\newcommand{\eea}{\end{eqnarray}}
\newcommand{\beq}{\begin{equation}}
\newcommand{\eeq}{\end{equation}}
\newcommand{\beqa}{\begin{eqnarray}}
\newcommand{\eeqa}{\end{eqnarray}}

\newcommand{\vev}[1]{\langle #1 \rangle}

\title{Entanglement String and   Spin Liquid with   Holographic Duality }
 \author[a]{Eunseok Oh,}
\author[a]{and Sang-Jin Sin}
\emailAdd{lspk.lpg@gmail.com}
\emailAdd{sangjin.sin@gmail.com}
\affiliation[a]{ Department of Physics, Hanyang University, Seoul 133-791, Korea }
\abstract{
We show that the quantum entanglement can be transmuted to a force using the holographic duality. 
First, we prove that there is an open string in the spectrum of the holographic fermion coupled with scalar. 
The string ends at two  fermions and its tension vanishes in the limit of zero scalar condensation.
We associate such string with a dimer and identify the scalar condensation as the degree of the dimerization. Together with divergently large entanglement entropy, the model is  expected to describe the Spin Liquid. As a consistency check, we show that  there is  a Mott transition as the dimerization proceeds. 
We suggest that  the string may  be observed in an ARPES experiment of spin liquid or clean Dirac material as a tower of  bands.
}

\subheader{
}

\arxivnumber{}

\begin{document}

\maketitle

\section
{\bf Introduction}  
Strong correlation gives many unexpected phenomena  which are very   interesting   as well as useful. 
Mott insulator, Strange metal and High Tc superconductivity are some of examples. Most interesting observation is the unreasonably fast equilibration both in  strange metal\cite{Sachdev:2011mz} and quark gluon plasma\cite{adams2005experimental}, which seems to request non-local force.  In such system, the low energy   spectrum  is  very different from that of high energy: QCD  string\cite{Veneziano:1968yb,Mandelstam:1974pi}   and fractionalized degrees in spin liquid \cite{ANDERSON1196,kivelson1987topology,Fulde:2007vx} are examples. Finding  such excitation often gives us a  way to analyze a very difficult system in a simple way.

Recently, gravity dual description \cite{Maldacena:1997re,Witten:1998qj} attracted much attention as a tool   \cite{Zaanen:2015oix,Hartnoll:2016apf} for  strongly interacting system (SIS). The basic idea is to utilize the universality  near the quantum critical point  and its similarity with a black hole which shares the same scaling symmetry and the temperature. 
Since the entropy of black hole is given by its  area \cite{Bekenstein:1974ax} rather than the volume, we have a holographic relation \cite{tHooft:1993dmi,Susskind:1994vu}  whose origin can be attributed to  the  equivalence principle: infinitely large number  of metrics can be identified. From the SIS point of view the holographic relation comes because the configurations of SIS at different energy scales are glued together \cite{VanRaamsdonk:2010pw,Maldacena:2013xja,Faulkner:2013ica,Hartman:2013mia,Oh:2017pkr} by  quantum entanglement  to form  one higher dimensional system. 
 It   is   very  intriguing  to ask  if there is a mechanism  by which  entanglements can give a sort of force. 
   
Motivated with these questions,  we studied the   fermionic particle  with  a scalar coupling in the dual gravity setup.  We found an exact two-point function   and  found that mass spectrum  is that of open string excitation. Its string tension  is  given by the  scalar condensation, therefore the string is tensionless in the zero condensation limit. 
The most intriguing part here is the origin of this stringy spectrum. 
For the hadronic physics, its appearance can be easily attributed to  the QCD string, however, it is rather mysterious in the condensed matter because  no confinement physics seems to be involved.
Nevertheless, the appearance of massive string spectrum indicates that qualitatively similar physics is in action. 

The open string should connect two fermions in the boundary which are the only object in our system,  to form a dimer, the spin singlet bound state.  In the absence of the scalar condensation, the string is massless   so that there is no cost in forming a dimer between the far separated fermions. 
However, in its presence, the string  has tension so that it should connect nearest two fermions to minimize its  energy.  In this picture, we suggest to call  the new string as `{\it entanglement string}',
  and  the scalar condensation  measures the degree of  the dimerization of the system, equivalently the string condensation. 

There can be a few possibilities with the condensation of the dimers: for example valence bond solid (VBS) 
or resonating valence bond (RVB) \cite{ANDERSON1196}, because spin singlet condensation does not form any magnetization. See  ref. \cite{Balents:2010tw,Sachdev:2011mz} for review and   references. 
The key is the  size of the entanglement entropy.  
Sometimes ago,  Ryu and Takayanagi \cite{Ryu:2006bv} showed that the presence of dual gravity request  infinitely large entanglement entropy. The only system with dimerization with such large entanglement entropy is the RVB state whose fundamental aspect  is  also the presence of high degree of entanglement entropy    \cite{savary2016quantum}.  
Therefore it is natural to identify our system as the spin liquid rather than spin solid. 
 
 As a consistency check, we investigated the metal insulator transition, since the spin liquid is known to be an insulator.  Can our   holographic model show such transition?  
 We will show that our  system  indeed has a  Mott  transition  as we increase 
 one or more of following three parameters:   scalar condensation, temperature, chemical potential. 
  
  Finally  we also discuss    a few ways to observe the entanglement string, 
  whose  direct experimental test   is the observation of the stringy spectrum.

\section{Fermion in   $AdS_{4}$ and  string at the boundary}
 In this paper we study the fermion dynamics   with a real scalar $\Phi$ so that the fermion action  is given by 
\begin{eqnarray} 
S_D=i\int d^{d}x\sqrt{-g}\bar{\psi}(\Gamma^M\mathcal{D}_M-m-\Phi)\psi +S_{bd}, \label{action}
\end{eqnarray}
 where  ${\cal D}_M = \partial_{M} +\frac{1}{4} \omega_{abM}\Gamma^{ab} -i q A_M $
is  the covariant derivative in the  asymptotically $AdS_4$ of radius $L$   whose explicit form is given in the appendix. $\Phi$ is a scalar field  whose dynamics is given by a free real scalar action with   
    mass $m_{\Phi}^2 L^{2}=\Delta(\Delta-3)=-2$ in the same gravity background.
If  $\Phi^{(0)}$ is a scalar field that  couples with the boundary fermion bilinear $\bar{\chi}\chi$  with dimension $\Delta=2$, it is embedded in the bulk field $\Phi$  such that  near boundary behavior of the latter is 
\be
\Phi(x,r)=\frac{\Phi^{(0)}}{r}+ \frac{M}{r^2} +\cdots.   \label{Phi}
\ee  
 Notice that in the zero temperature zero chemical potential limit, {\bf the first two terms are two independent exact solutions} of the equation of the scalar field equation in the probe limit.  Since we are looking for spontaneously generated gap at zero temperature, we set the source   
 $\Phi^{(0)}=0$. Notice that the dimension of $M$ is that of  mass squared.  
In the appendix, it is shown that the components of the   Green function of fermions in the boundary  theory  are given by 
\begin{equation} 
G^{R}_{11}(\omega, k) = \frac{(\omega-k_{1} )\Gamma(\frac{1}{2}-m)\Gamma(m+\frac{1}{2}+\frac{k^2-\omega^2}{4M})}{2M^{\frac{1}{2}-m} \Gamma(\frac{1}{2}+m)\Gamma(1+\frac{k^2-\omega^2}{4M})}, \quad  G^{R}_{22}=G^{R}_{11}(\omega, -k) , 
\label{preG22}
\end{equation} 
and $G^{R}_{12}=G^{R}_{21}=0$  in   the standard quantization.
 It can be easily checked that in the limit of $M\to 0$, 
  our result is reduced to that in ref.  \cite{Iqbal:2009fd}.
The spectral function is given by the imaginary part of the Green function, which can be easily recovered from Kramers-Kronig relation or  by the   prescription   $\omega\to \omega+ i0^{+}$. 

What is remarkable in this result is that the singularities of the Green function   are given by those of the Gamma function which has poles at all non-positive integer so that 
the spectrum of the theory is 
\be
\frac{k^2-\omega^2}{4M} +m+\frac{1}{2}=-n, \quad {\rm for} \quad n=0,1,2,\cdots, \label{Regge}
\ee
 which is nothing but the spectrum of the relativistic open string 
 \be 
 \alpha' m_{n}^{2}=  (n+m+\frac12 ), 
 \ee
  whose string tension
 $T=1/(2\pi \alpha') $ 
with   $\alpha'=\frac1{4M}$.  Eq. (\ref{Regge}) is the Regge trajectory with slope $\alpha'$ and intercept $\alpha_{0}=-m-1/2$. 
This is an analogous situation of the discovery of the string from the  Veneziano amplitude\cite{Veneziano:1968yb}, where the origin of the string could  be attributed to the confined color flux tube later. 
   In $M\to 0$ limit, that is, in the tensionless limit,  
    the whole tower of string spectrum is reduced to that of a massless particle. 
   Notice that    the lowest spectrum  given by $n=0$
   \be
   m_{0}^{2} =4M(m+\frac12 ),\label{lowest}
   \ee 
shows  that even for  $M\neq 0$,   we can still have the massless  spectrum  if $m=-1/2$.   
 The same formula also  shows that we get  tachyonic spectra for $m<-1/2$. In this paper we consider only   $-1/2\leq m \leq 0$. 
 
  \subsection{Entanglement string and dimers}
\begin{figure}[ht!]
\centering
       \subfigure[VBS]
    {\includegraphics[width=4 cm]{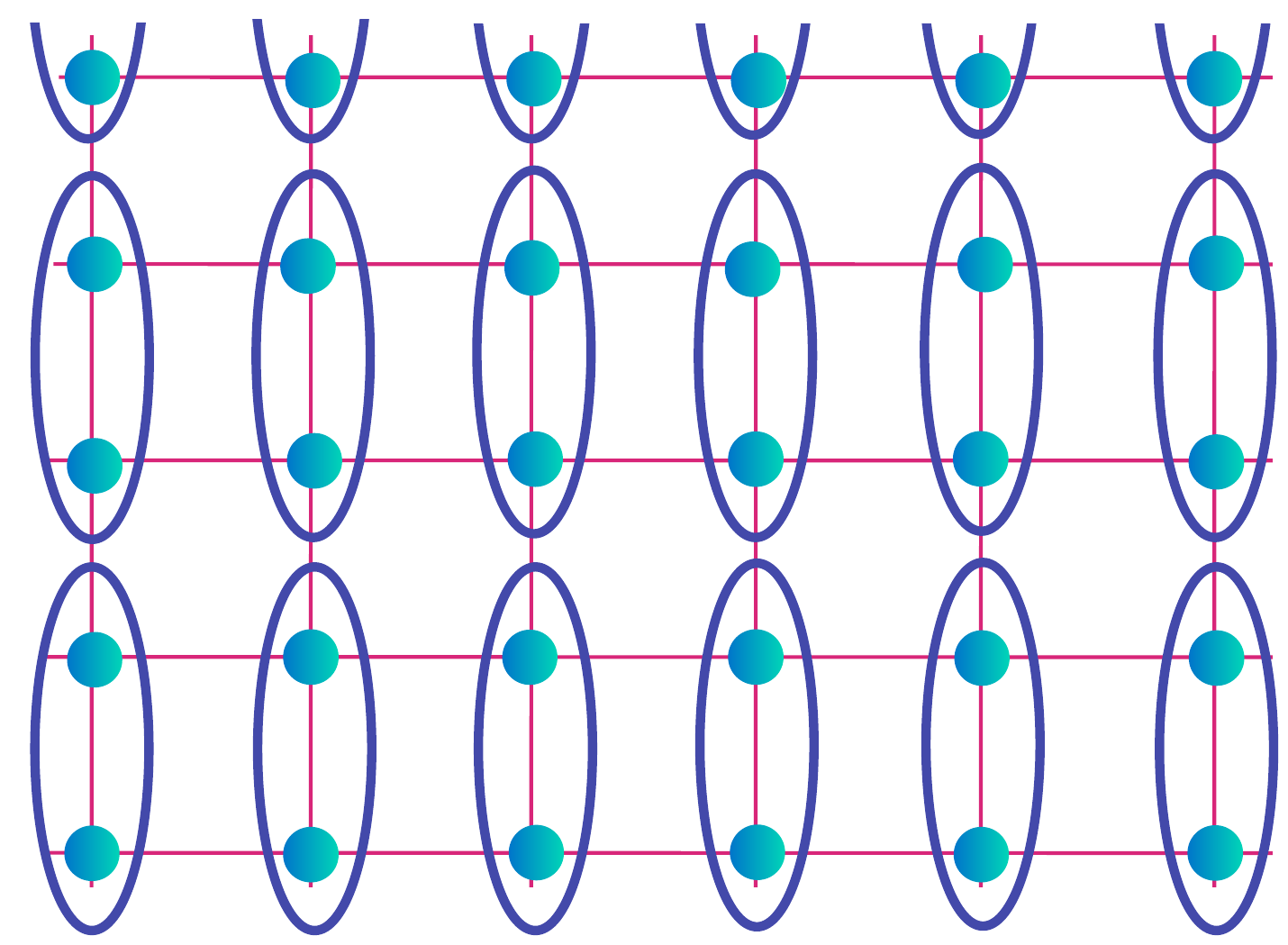}  }
        \hskip 1.5cm
           \subfigure[RVB]
    {\includegraphics[width=4.7cm]{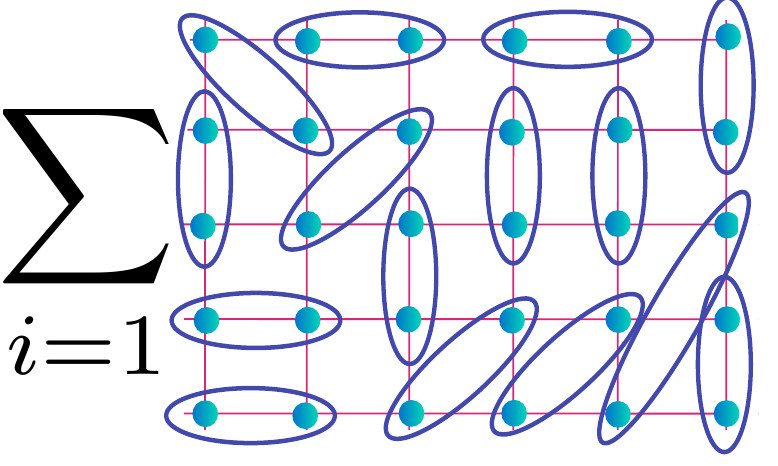}  }
          \caption{ (a) Valence bond solid phase: The rotation symmetry is broken 
          and can not be described by a scalar condensation.  
          (b) The rotation symmetry is restored and huge entanglement entropy is guaranteed by the superposition of all possible dimer condensed states. Such system is possible holographically without lattice. 
              The figures are minor adaptation from  ref. \cite{Sachdev:2012dq,Sachdev:2015slk}.} 
           \label{fig:VBSRVB} 
\end{figure}

The most intriguing question here is about the origin of the string. 
In fact, the presence of tower of spectrum has been appeared   in the numerical plots of earlier literature, although its meaning has not been discussed.  
From the dual gravity point of view,  it is a very special type of Kaluza-Klein (KK) spectrum of the bulk 
fermion. Usually the KK spectrum is quadratic in $n$, $m_{n}^{2}=c.n^{2}$, because one of the momentum component is quantized. Here it is linear:  the AdS has  a character of box drived by the quadratic gravitational potential. Such geometric picture gives an intuition why we have Regge trajectory, nevertheless, it  does not help to understand its  origin from the  view  of the original strongly interacting fermions. 
Let's collect some relevant facts and  proceed to identify the origin of the string in a few steps. 
\begin{enumerate}
\item We assumed  the duality between the original fermion $\chi$ and the bulk fermion $\psi$  in AdS$_{4}$. 
One should notice that the spectral function is that of  $\chi$, and so is  the string.  
 The original  fermion  lost its character as an elementary excitation due to the interaction
and became a D0 brane\cite{Polchinski:1996fm}, an object whose dynamics is given by an open string attached to it. 
Summarizing, the fermion particle in AdS is the holographic   image of the string in the strongly interacting fermion system at the boundary. See  the figure \ref{fig:entstring}(a).
\item 
What is the physical reality of the string   in the original   system?  
Since the open string should end somewhere,  our string should connect two fermions. 
Such  fermion-string-fermion system can be identified as a dimer, which is a spin singlet Einstein-Podolski-Rosen (EPR) pair that is used as  a building block of RVB (resonating valence bond) state \cite{ANDERSON1196,kivelson1987topology}. 
See figure \ref{fig:entstring}(b). 
Also it is natural to interpret the $\Phi$ as the low energy description of the dimer where the separation of the two fermions can be neglected. Then, the condensation of the scalar should describe the 
degree of dimerization of the system.  
\begin{figure}[ht!]
\centering
 \subfigure[dual string]
    {\includegraphics[width=2.5cm]{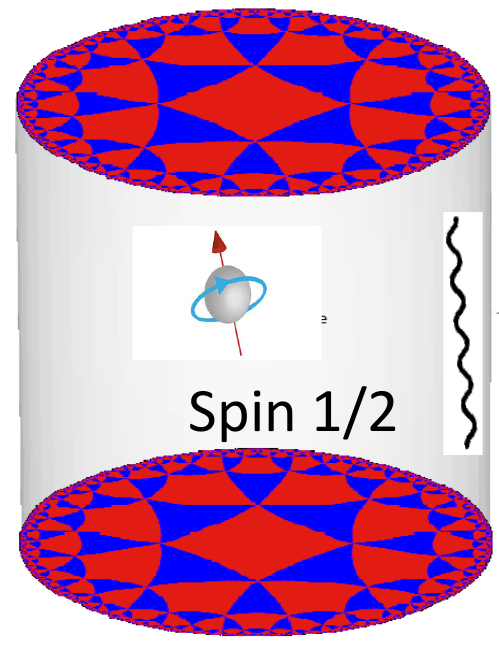}  }
    \hskip 0.5cm
    \subfigure[dimer and open string]
    {\includegraphics[width=6cm]{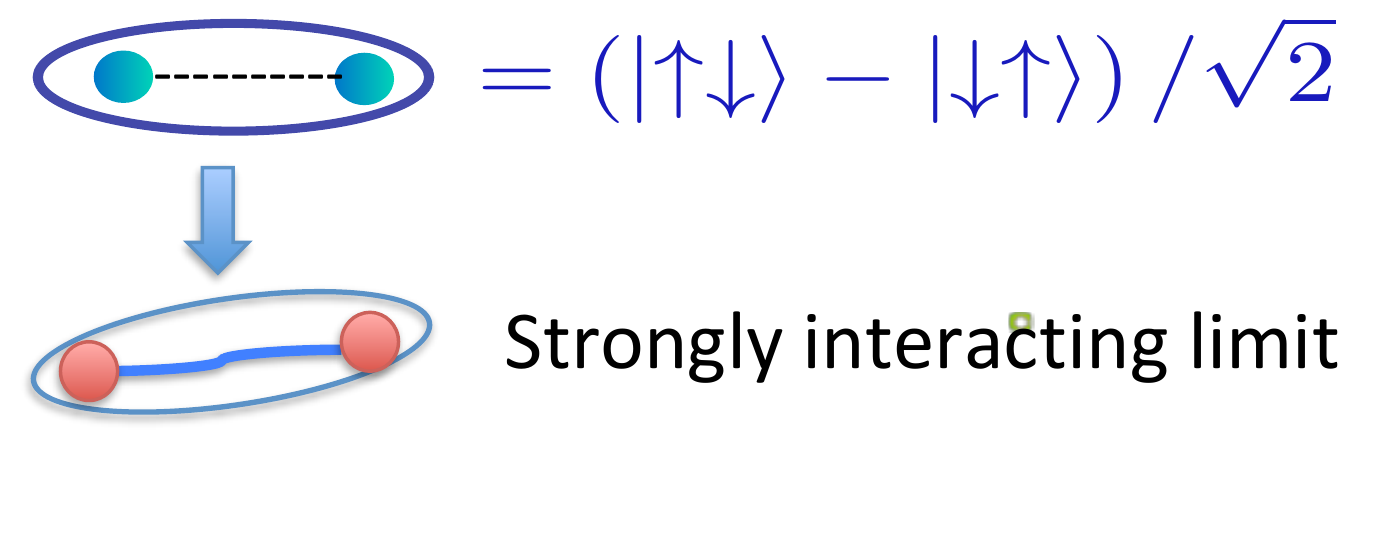}  }
       \subfigure[entanglement string]
    {\includegraphics[width=5cm]{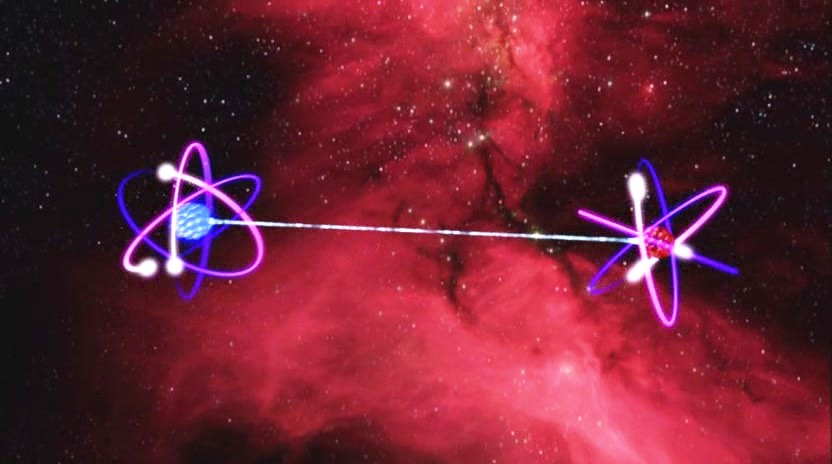}  }
          \caption{
          (a) An AdS fermion is dual to the string at boundary.  
          The figure is an adaptation  from  the Wikipedia for `AdS/CFT correspondence'. 
          (b) A dimer is a spin singlet combination of  two fermions.                 
               In the strong interaction limit, the  particle nature of the fermion (blue dot) 
             seems to be  transformed into D0-particle (red dot)  so that
              its dynamics is given by an open string.                             
              (c) The cartoon for quantum entanglement  (from \cite{Arjun:2016}), where 
            a string is connecting two entangled fermions, can be a physical reality. 
          In the absence of the scalar condensation, the string  can connect two entangled fermions of any distance without energy cost.     } 
           \label{fig:entstring} 
\end{figure}
\item In the limit of zero scalar condensation $M\to 0$, namely when there is no other dimers around the one we consider, the string tension is zero and there is no energy cost in connecting arbitrarily  far separated  EPR pair. 
Therefore it is natural to call  our string  as {\bf entanglement string}. 
See figure \ref{fig:entstring}(c). 
As scalar condensation $M$ increases, 
    the dimers are more and more tightly bound by the string tension so that  the system certainly  prefers dimers formed between nearest fermions to minimize the energy. 
Long strings should be suppressed for such case.
 \item   ${\bar\chi}\chi({\bf x})$ is a spin singlet and charge neutral regardless of   $\chi$ being Dirac or Majorana fermion. Since the open string is connecting two fermions  not fermion anti-fermion, 
they are better to be Majorana fermions or a spinons \cite{ANDERSON1196,kivelson1987topology}.  
 This also confirms that the condensation of the scalar can  encode only the degree of dimerization. 
 \item Half of the origin of the stringy spectrum is  the presence of the dual space itself, because fermion in a AdS space   has Kaluza-Klein spectrum.  The other half is the specific power 2 in the  Eq. (\ref{Phi}).
 It depends on the  spacetime dimension $d$, the spin of the condensed field $s$ and the conformal dimension $\Delta$ of the operator the bulk field is coupled to.  Namely the string spectrum is the consequence of  $ \Delta-s=2$. For the fermion bilinear, $\Delta=d-1$. 
\item    There still are a few options for configuration of dimers: either it can have some order in their alignment  called valence bond solid ({\bf VBS}), or it can be  symmetry unbroken state called resonating valence bond ({\bf RVB}). 
Their  main difference is the amount of the entanglement entropy  \cite{savary2016quantum}.  
The VBS can not be described by a scalar condensation because the rotation symmetry is broken.  
See figure \ref{fig:VBSRVB}(a,b). 
According to Ryu-Takayanagi\cite{Ryu:2006ef}, any  system with holographic description   has divergently large entanglement entropy proportional to the interface area. Therefore we can  identify the system of holographic fermions  with  scalar condensation as {\bf spin liquid} or resonating valence bond ({RVB}), the system of dimer condensation with maximal entanglement. See figure \ref{fig:VBSRVB}(b). 
The identification is consistent both symmetry and entanglement entropic point of views. 
  \end{enumerate}
The message is that  the presence of the holography which by itself is a consequence of entanglement, 
  gives a set up where the quantum entanglement (QE) acts on a system as a force, the string tension, enforcing the system to realize the dimerization by nearest fermions.  
 It also suggests that even without a strong correlation, a string   is associated with EPR pair
 which resembles the statement `{\bf ER=EPR}'
 \cite{Maldacena:2013xja,Jensen:2013ora,Seki:2014pca}. 
This may be useful to explain the non-local forces that seem to be responsible for the rapid thermalization   of the  strange metal \cite{Sachdev:2011mz} and quark gluon plasma \cite{adams2005experimental}.

  \subsection{More on experimental implementation}
\begin{figure}[ht!]
\centering
   \subfigure[gapless $m=-\frac12, M=1$]
    {\includegraphics[width=4cm]{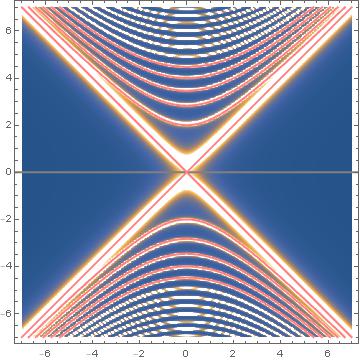}  }
    \hskip 1.5cm
    \subfigure[gaped   $m=-\frac14, M=1$]
    {\includegraphics[width=4cm]{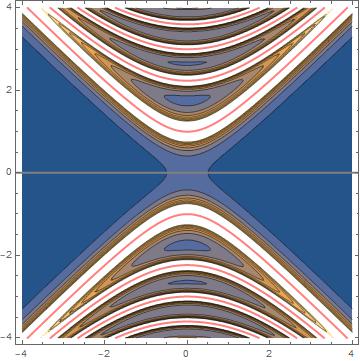}  }
         \caption{Stringy spectrums (a) without and (b) with gap. White peaks are contour plots of the numerical results and red lines are plots of the analytical result Eq. (\ref{Regge}).  Finite widths of white peaks are effects due to the interaction and small but finite temperature.  }  \label{fig:stringyspec}
\end{figure}

 The most direct prediction of the theory is the existence of the string. 
    Finding string means observing its spectrum given in figure \ref{fig:stringyspec}.
    We expect that such tower of string spectrum may   be  found  in the ARPES experiment of some Dirac materials with strong correlation.  

   Any Dirac material with small fermi surface, which is realized when the material is clean enough, 
   is strongly interacting. Therefore we expect that the stringy spectrum can be observed  in the experiment with clean Dirac material. Notice that  the higher spectral tower  with large $n$ 
    is suppressed only by $1/\pi n$  instead of $1/n!$, which is naively expected from the Gamma function residue. This can be seen easily from Eq. (\ref{Regge}) using the formula $\Gamma(1+z)\Gamma(1-z)=\pi z/\sin\pi z$.  The role of another Gamma function in the denominator of  Eq. (\ref{Regge}) is crucial for enhancing the spectral weight of higher order spectrum. 
     Using a non-Dirac spin liquid material  will also give a chance to observe it, because the  difference between different quantum critical points is   only quantitative one.       
    
    By figuring the elementary excitation as a dimer we may visualize how we can observe a stringy tower of spectrum more easily. 
When we shine a photon into a dimer, one of the end point fermion is affected if it is Dirac particle and it excites the string between two fermions. Such excitation will be detected in ARPES experiment. If we consider the end point fermions as  
 neutral Majorana particles which exist due to the spin charge separation, similar stringy spectrum may be observed in the susceptibility measurement of the neutron scattering with  spin liquid target material. 
 We expect that there should be maximum of $M$ corresponding to the saturation of dimerization. 
     
     Finding such stringy spectrum  would be extremely interesting, since it  is an evidence of  the holographic dual space  as well as  the entanglement string in real matter. 

\section{Mott transition}
   Now  lets do some consistency check.  The spin liquid is  known to be a Mott insulator. 
    Is our system a Mott insulator for full dimerization or equivalently  for large enough scalar condensation $M$?  In this section, we will see  that  in fact   our system has Mott transition  as we increase the scalar    condensation. 
    
\begin{figure}[ht!]
\centering
   \subfigure[Three phases]
    {\includegraphics[width=5cm]{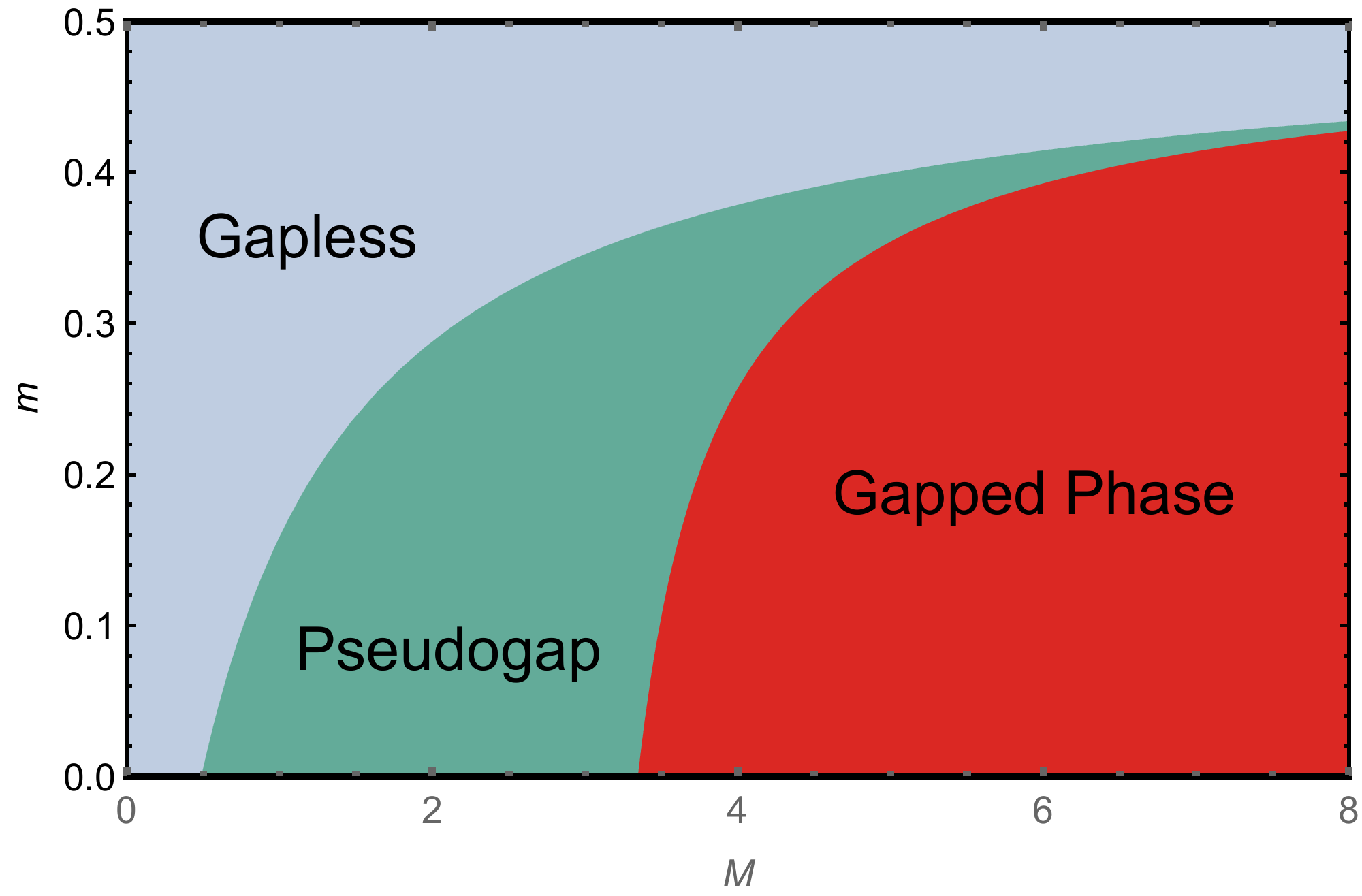}  }
    \hskip 1cm
        \subfigure[Subregions of Gapless Phase]
    {\includegraphics[width=5cm]{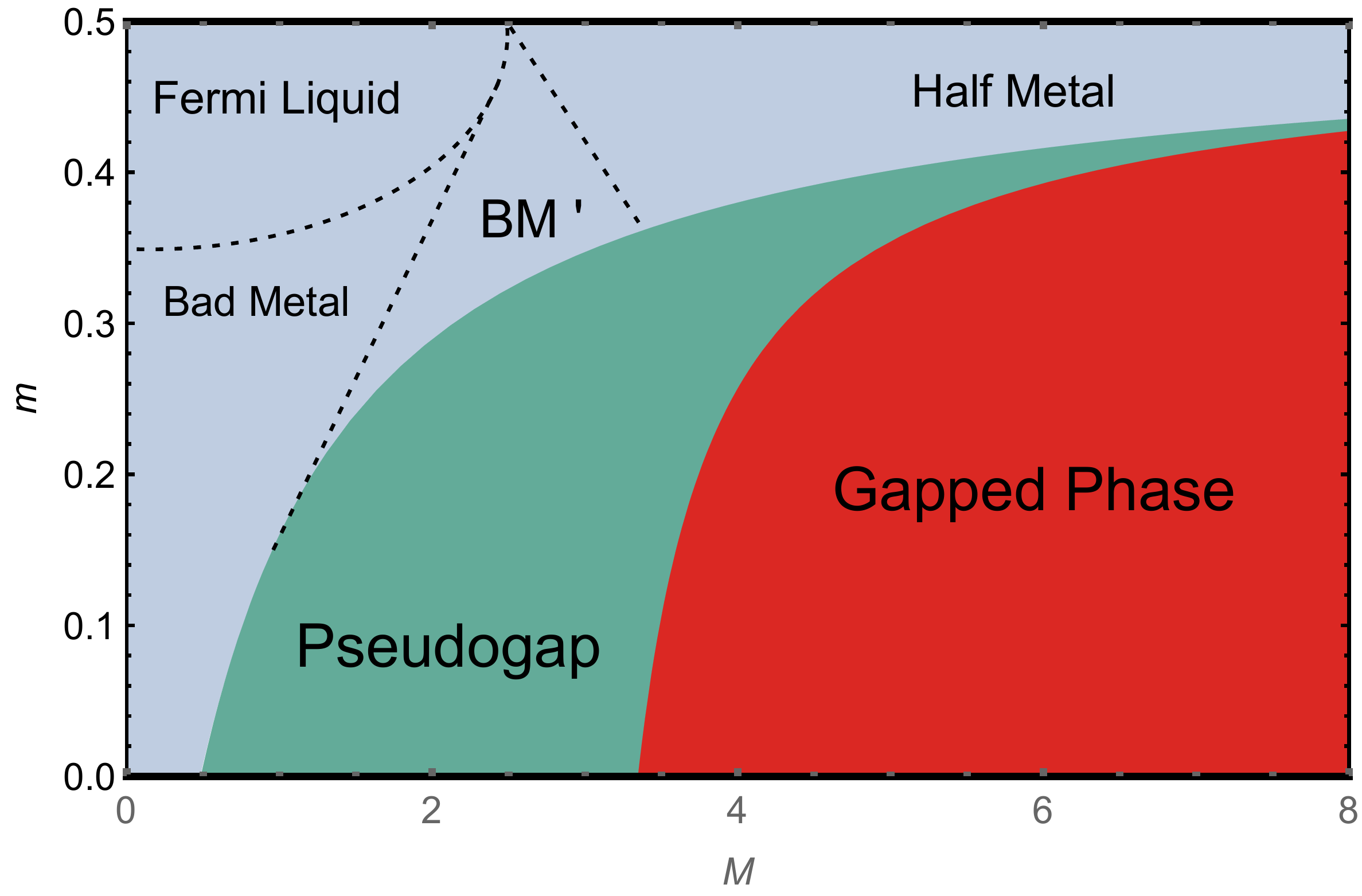}  }
         \caption{$m$-$M$ Phase Digrams.  
         In all phase boundaries, the changes are smooth crossover. Notice that for all values of $m$, 
         there is a transition from gapless to gapful phase through the pseudogap regime. 
         Notice that  m in the vertical axis   means its absolute value $|m|$. 
         }  \label{fig:phase1}
\end{figure}

    The study of the spectral function shows that  there are three easily distinguishable phases: gapless, pseudo-gap and gapped phases as one can see in the 
    $(M,m)$ phase space   diagram given  in figure  \ref{fig:phase1}. 
Here, gapless means the existence of degrees of freedom at the Fermi sea $\omega=0$ and  
gapped phase does not have any, pseudo gap means the depletion of DOS at the central peak.  
We choose the onset of pseudo gap as the 10\% depletion.

We remark that  a similar model  with  the Pauli interaction instead of scalar coupling  \cite{Seo:2018hrc}   it was shown that the model exhibits the Mott transition  \cite{MOTT:1968aa}    connecting the free fermion point\cite{Cubrovic:2009ye,Cubrovic:2010bf,Medvedyeva:2013rpa}  and gapped phase  \cite{Edalati:2010ge,Edalati:2010ww}  under increasing the coupling  strength. 
To compare with our phase diagram with that of the Pauli interaction, we divide the gapless region into 4 subclasses: Fermi Liquid like (FL), Bad Metal (BM), Bad Metal prime (BM') and half Metal (hM). 
The characters of these phases are qualitatively similar to those of Pauli case and we refer the interested readers to the ref. \cite{Seo:2018hrc}. 
From the fact that $m=-1/2, M=0$ is the free fermion point and there is a large gapped phase region, any line connecting these two can realize the Mott transition. Furthermore, increasing $M$ induces the transfer of the degree of freedom from the central peak to shoulder peak. 
These features are common  with the model with Pauli interaction   studied  in  \cite{Seo:2018hrc}.  

There is an important difference, however: while 
  there is a Mott transition  for any value of $m $ as  $M$ varies  for the scalar interaction, 
there is none for fixed $m$ if $|m| >0.35$ in the Pauli interaction case.  

\begin{figure}[ht!]
	\centering
 \subfigure[m-M PD for various T]
    {\includegraphics[width=4.7cm]{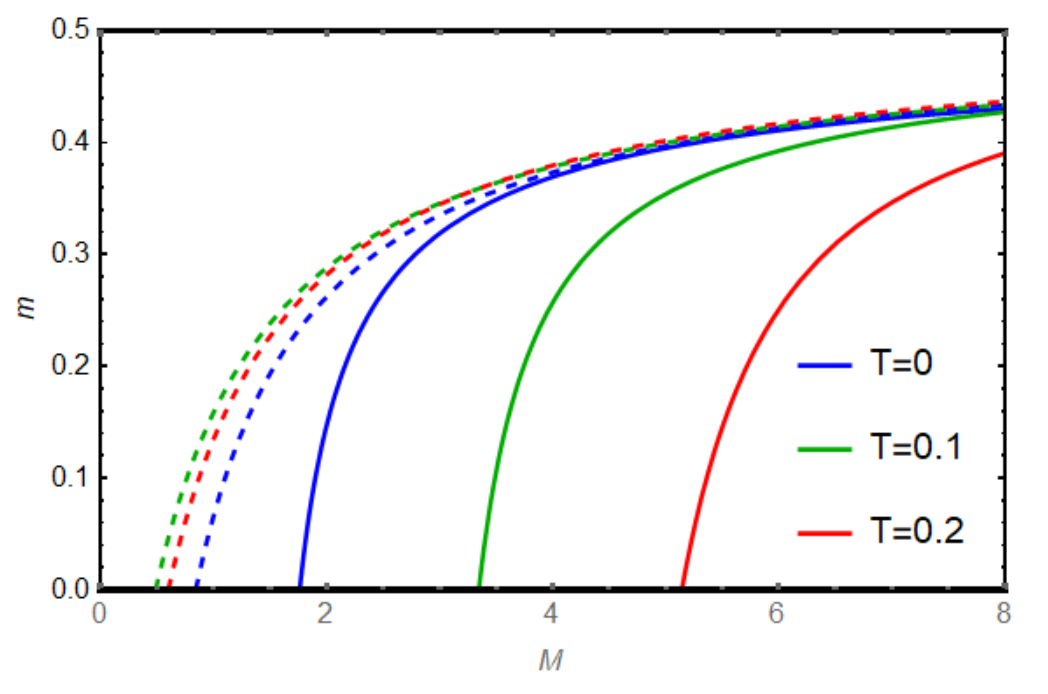}  }
 \subfigure[m-M  PD for various $\mu$]
     {\includegraphics[width=4.8cm]{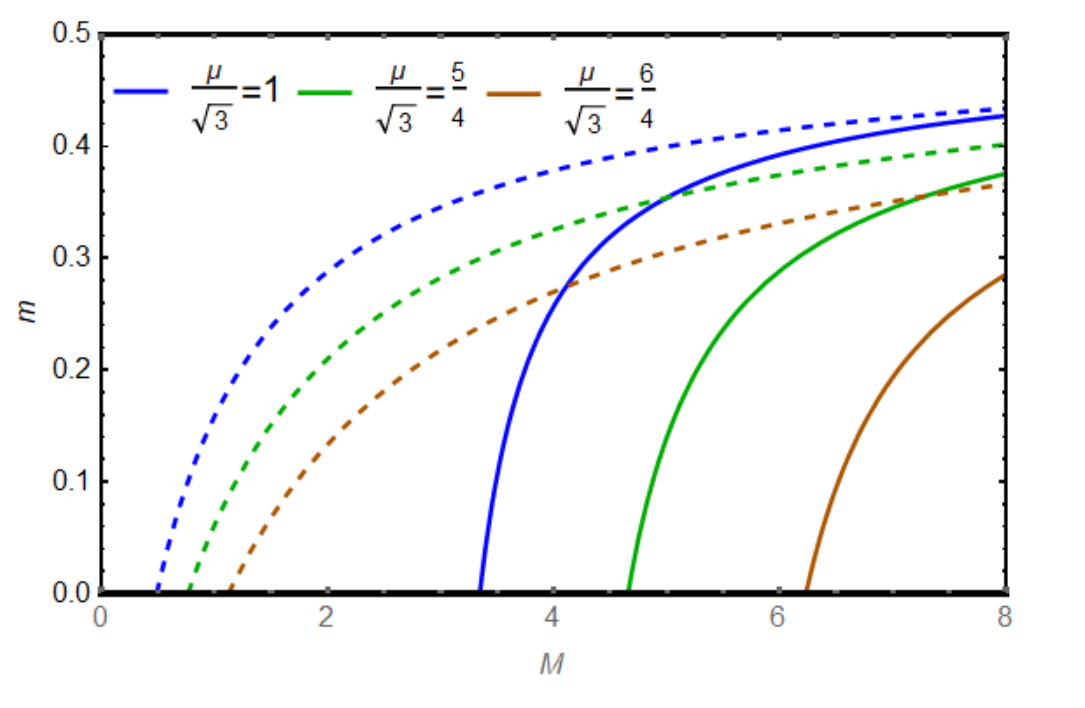}  }
 \subfigure[$\mu$-$T$  PD ]
     {\includegraphics[width=4.7cm]{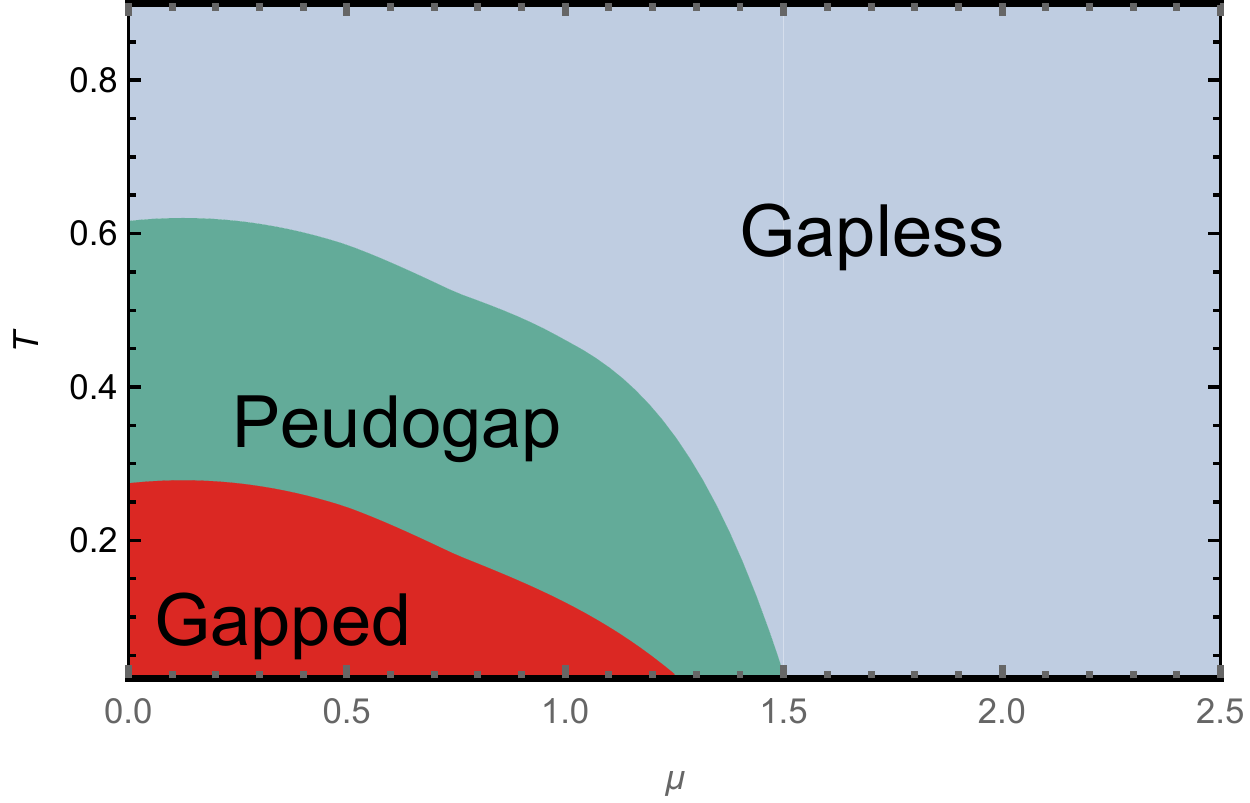}  }    
     \caption{$|m|$-$M$ Phase Digram (PD) for (a) T-evolution (b) $\mu$-evolution.
     Dotted lines are the phase boundaries for Gapless-PG phases and real lines are those for PG-Gapped phases.
     The Phase boundaries move to the higher $T$ or $\mu$ region as we increase  $T$ or $\mu$. 
     (c)  $\mu$-$T$ Phase Diagram. Here $m=-0.3$ and $M=4$. } 
       \label{fig:mMmuT}
\end{figure}

Now we study the  temperature and chemical potential  dependence of the phase diagram. 
 The figure \ref{fig:mMmuT}(a,b) show how the $m$-$M$ phase diagram(PD) changes as we change those parameters respectively. We can easily sea that as we increase temperature,  the boundary of PG and Gapped phases  is moving to the higher $M$ region significantly, while the boundary of Gapless and PG phase  does not move much. The chemical potential evolution shows similar behavior qualitatively. 
 Phase boundaries move to the higher $\mu$ region as we increase $\mu$. These are consistent with  our expectations that Gap formation  is harder  at higher temperature or higher chemical potentials. 
We   also draw  $\mu$-$T$ phase diagram  explicit in figure \ref{fig:mMmuT}(c). It would be interesting if these diagrams can be compared with experimental data. 

\section{Summary and Discussion} 

In this paper we  investigated  the scalar coupling of the fermion.  
First, it is shown that two entangled fermions are connected by a  string 
and the string tension is proportional to  the scalar condensation, therefore for  vanishing condensation, the string becomes tensionless and all  spectrum  becomes massless
 so that the string is invisible.    
Secondly,  the scalar condensation is identified as the dimerization and the system is suggested to be describing the spin liquid.
 Thirdly, we found that there is a Mott transition  as we change the scalar coupling.
 
   The scalar field condensation   in this paper is just  one possible mechanism of gap generation and it is identified as dimerization of the system. It would be   interesting if we can similarly identify other gap generating interactions  as   physical phenomena.  In this regard, we mention that 
in ref. \cite{Seo:2018hrc}, we suggested that  when the mass generation is associated with    the  Pauli  term, such instability can be related to the ``nesting'' for the charge density wave. This means that the gap generated by the Pauli term is associated  with the CDW order just as the gap generated by the   Yukawa term is associated with  dimerization.   

We   expect that  for all gap generating mechanism, similar trajectory should appear although it may not be strictly a linear one. Also, notice that in this paper the scalar field is considered as a probe. It would be interesting to couple it with the gravity  so that the condensation  is determined  including the gravity back-reaction.

   The appearance of the stringy spectrum can be more universal than we expect because  matter can become a SIS  in many ways: first, by having a small Fermi surface \cite{Seo:2016vks,Seo:2017oyh,Seo:2017yux} as in the Dirac materials which  form rather wide class, second by having small Fermi velocity   as in the case of all transition metal oxides. This is because the effective coupling inside matter is 
$\sim  {\alpha} /{\epsilon v_{F}}$, where $\epsilon$ is a screening constant. 
We wish that the entanglement string can be found in the spectrum of diverse material.

 \acknowledgments
 This  work is supported by Mid-career Researcher Program through the National Research Foundation of Korea grant No. NRF-2016R1A2B3007687. 

\bibliographystyle{JHEP}
 \bibliography{Refs_scalar.bib}
 
\appendix
\section{Femion spectral function}
The basic idea of fermion correlation function is that we take the dual of the fundamental fermion as the bulk fermion whose action  is given by Eq. (\ref{action})
 in the bulk background given by RN-$AdS_4$
\begin{equation}
\begin{split}
d s^2=-\frac{r^2}{L^2}f(r)d t^2+\frac{L^2}{r^2f(r)}d r^2+\frac{r^2}{L^2}d x^2, \quad
f(r)=1-\frac{r_+}{r^3}+\frac{r_+ \mu^2}{r^3}+\frac{r_+^2 \mu^2}{r^4} .
\end{split}
\end{equation}
Gauge potential is given by $A=\mu(1-\frac{r_+}{r})dt$ and the horizon of this metric $r_+=\frac{1}{3}(2 \pi T + \sqrt{4\pi^2 T^2+3 \mu^2})$.
In this paper, we treat both the scalar $\Phi$ and and the fermion $\psi$  
as probes to get analytic solution.  Near the boundary, the real scalar $\Phi$ has usual expansion  given by 
\be
\Phi=\frac{M_{0}}{r^1} + \frac{M}{r^2} \cdots .
\ee
To consider the scalar condensation that arizes spontaneously, we set the souce $M_{0}=0$ with positive condensation   $M>0$.  Notice that  $M=\vev{ \bar{\chi}\chi}$  has dimension 2, since the boundary   fermion $\chi$ has the dimension 1 in 2+1. 
\footnote{Notice that  similar conclusion can be drawn for Majorana fermion $\chi$.  
In that case, one should replace the mass term to the Majorana one. }
 We choose Dirac matrices as follows: 
\begin{equation}
\Gamma^r={\begin{pmatrix}
	1 &  0 
	\\
	0 & -1 
	\end{pmatrix}},
\Gamma^t={\begin{pmatrix}
	0 & i \sigma_2 
	\\
	i \sigma_2 & 0 
\end{pmatrix}},
\Gamma^x={\begin{pmatrix}
	0 &  \sigma_1 
	\\
	 \sigma_1 & 0 
	\end{pmatrix}},
\Gamma^y={\begin{pmatrix}
	0 &  \sigma_3 
	\\
	\sigma_3 & 0 
	\end{pmatrix}}.
\end{equation}
Following \cite{Liu:2009dm}, we   Fourier transform followed by introducing  $\phi_{\pm}$  
\beqa
\psi_\pm = {(-gg^{rr})}^{-\frac{1}{4}} 
 \phi_\pm, \quad \phi_\pm = 
\left(
\begin{array}{ccc}
  y_\pm    \\
  z_\pm    \\
\end{array}
\right), \label{Eq:psi_pm}
\eeqa  
and  $\xi_\pm $ by 
$  \xi_+ =i \frac {y_-}{z_+}, \hbox{ and } \quad \xi_- = - i\frac{ z_-}{y_+}.
 $ 
Then the   equation for $\xi_{\pm}$ can be written as
\begin{equation}
\begin{split}
\label{floweq}
\sqrt{\frac{g_{ii}}{g_{rr}}}\left(\partial_r+2(m+\frac{M}{r^\alpha})\sqrt{g_{ii}}\right)\xi_{\pm}=\pm\left(k_1\pm\sqrt{\frac{g_{ii}}{g_{tt}}}(w+A_t)\right)\xi_{\pm}^2\mp\left(k_1\mp\sqrt{\frac{g_{ii}}{g_{tt}}}(w+A_t)\right).\nonumber
\end{split}
\end{equation}
 If we set $\mu\rightarrow0, T\rightarrow0$ and $\alpha=2$,  the   Eq.(\ref{floweq}) for $\xi_-$ is reduced
 to 
\begin{equation}
\label{deqxim}
\begin{split}
r^2\partial_r\xi_-+(2 m r + 2 \frac{M}{r})\xi_-+(k_1-w)\xi_-^2-k_1-w=0 .
\end{split}
\end{equation} 
The solution to Eq. (\ref{deqxim}) is given by $\xi_{-}=N/D, $
with 
\begin{align}
\label{solxim}
N=& \frac{C_0r^{2m}}{m-\frac{1}{2}} (k_1+w){}_1F_1(h_1;\frac{3}{2}-m;\frac{M}{r^2})   \\
&-\frac{4i e^{i	m\pi}M^{m+\frac{1}{2}}}{(k_1-w)}\left[((m+\frac{1}{2})r+\frac{M}{r}){}_1F_1(h_2;\frac{3}{2}+m;\frac{M}{r^2})
+\frac{k_1^2-w^2-4M}{4(\frac{3}{2}+m)r}{}_1F_1(h_2;\frac{5}{2}+m;\frac{M}{r^2})\right] , \nonumber\\
D=& 2C_0r^{1+2m}{}_1F_{1}(h_0;\frac{1}{2}-m;\frac{M}{r^2}) +ie^{im\pi}M^{\frac{1}{2}+m} {}_1F_{1}(h_2;\frac{3}{2}+m;\frac{M}{r^2}) , \\
&~~~\quad h_0=\frac{k_1^2-w^2}{4M},  \quad h_1=1+\frac{k_1^2-w^2}{4M}, \quad
h_2=m+\frac{1}{2}+\frac{k_1^2-w^2}{4M}.
\end{align} 
For  the retarded Green function $G_{R}= \lim_{{r\rightarrow\infty}} 
r^{2m}{\rm diag} (   	\xi_{+}, \xi_{-} )$, 
\begin{equation}
G_{22}= -\frac{i e^{im\pi}(1+2 m)M^{\frac{1}{2}+m}}{(k_1-w) C_0(k_1,w,M)} ,
  \quad {\rm if} \quad  m > -\frac{1}{2}
\end{equation}
can be determined with  $C_0$  determined by the in-falling condition  $\xi_{-}(r)|_{r\rightarrow0}=-i$ to be given by  Eq. (\ref{preG22}).

\end{document}